\documentclass[11pt,twoside,english]{article}
\usepackage{graphicx}

\usepackage{asp2010}

\begin{document}
\title{High performance astrophysics computing}
\author{R. Capuzzo Dolcetta, M. Arca Sedda, A. Mastrobuono Battisti, D. Punzo, M. Spera}
\affil{Dep. of Physics, ``Sapienza'', Universit\'a di Roma, Piazzale Aldo Moro 2, 00185 Roma, Italy}
\begin{abstract}
The application of high end computing to astrophysical problems, mainly
in the galactic environment, is developing for many years at
the Dep. of Physics of Sapienza Univ. of Roma. The main scientific
topic is the physics of self gravitating systems, whose specific subtopics
are: i) celestial mechanics and interplanetary probe transfers
in the solar system; ii) dynamics of globular clusters and of globular
cluster systems in their parent galaxies; iii) nuclear clusters formation
and evolution; iv) massive black hole formation and evolution; v)
young star cluster early evolution. In this poster we describe the
software and hardware computational resources available in our group
and how we are developing both software and hardware to reach the
scientific aims above itemized.
\end{abstract}

\section{Introduction}

Celestial mechanics is one of the most classic examples of chaos in physics: 
the mutual gravitating systems show a chaotic behaviour,
being extremely sensitive to differences in initial conditions. This
problem can be only partially controlled using high-order integration
algorithms. The intrinsic difficulty of the problem is summarized
by the so called double divergence of the pair interaction potential
$U_{ij} \propto 1/r_{ij}$, where $r_{ij}$ is the distance between particle $i$ and particle $j$.

The ``ultraviolet'' divergence corresponds to gravitational encounters
at vanishing distance, while the ``infrared'' divergence means that
the force never vanishes. 
The computational problems arising from these divergences make the
classic gravitational \emph{N}-body problem unique.

\section{The NBSymple code performance}

Here we present benchmark tests of our sympletic \emph{N}-body
code, NBSymple (Capuzzo-Dolcetta et al., 2011, New Astr., 16, 284)
running on hybrid CPU+Graphic Processing Units (GPUs) architectures.
Specifically we ran some benchmark on JAZZ, a hybrid CPU+GPU cluster
managed by CASPUR (see http://www.caspur.it/en/).

In Fig. 1 (left panel)  we report the relative speed up, $S_{n}=T_{p}(1)/T_{p}(n)$,
where $T_p(n)$ is the time spent using $n$ GPUs.
The approximately linear speed up of our code is evident (the slope
of the best fit is 0.97). 

\begin{figure}
{\includegraphics[width=6.5cm]{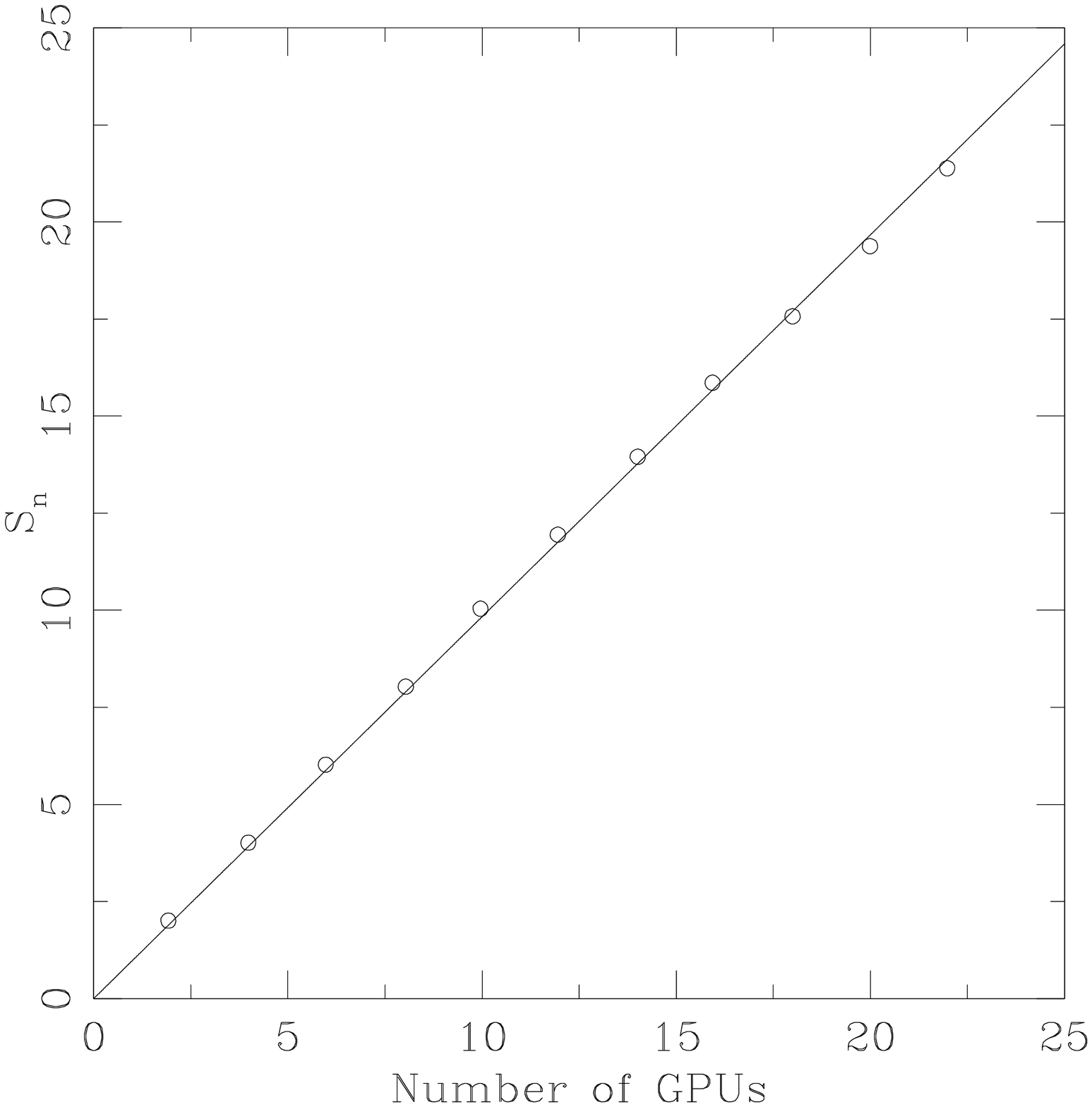}}
{\includegraphics[width=6.5cm]{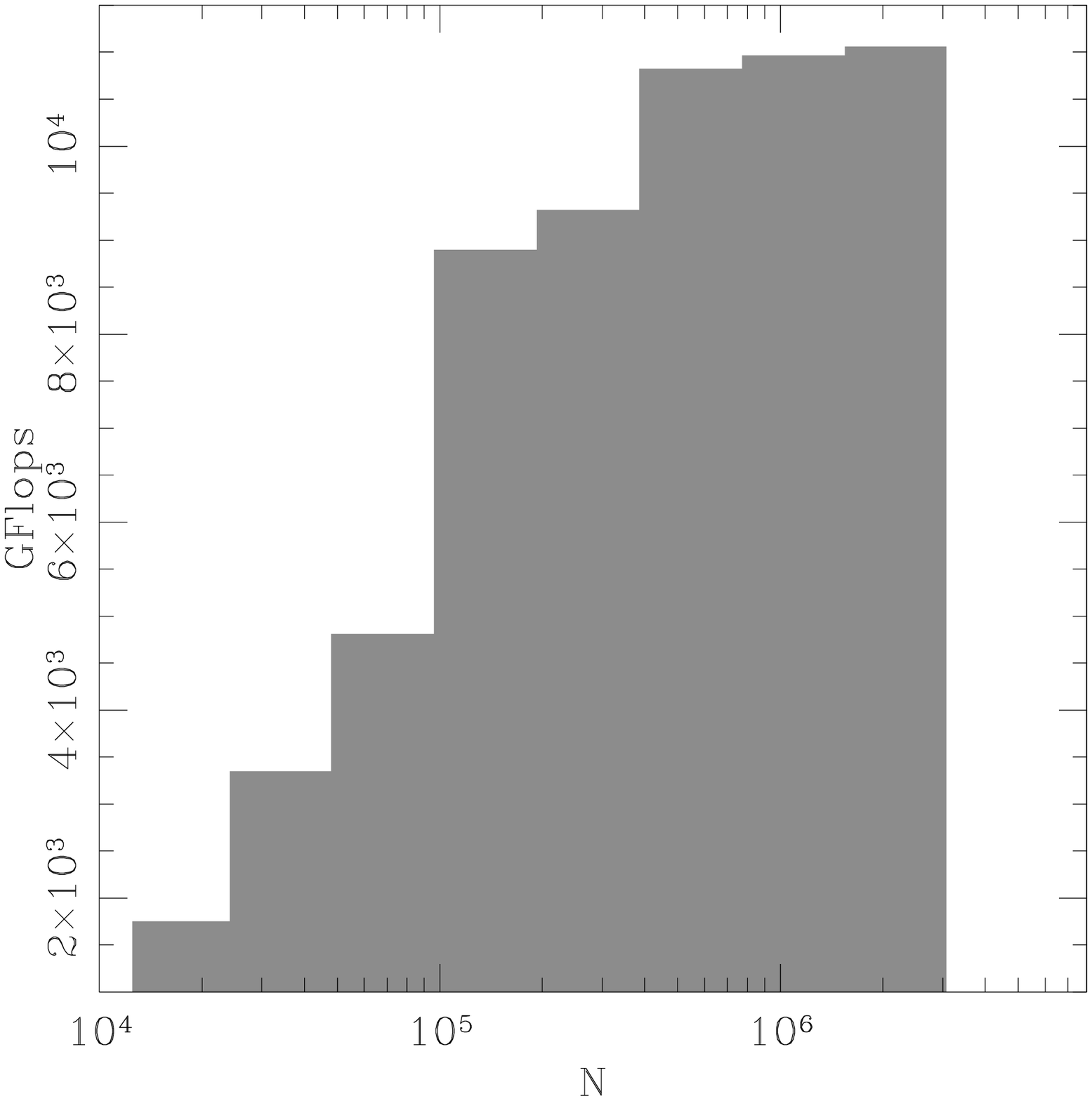}}
\caption{Left panel: the speedup in function of the
number of GPUs used. The number of particles is $N$=1,966,080.
Right panel: NBSymple performance in double-single precision, for different
values of $N$, using 11 nodes.}
\label{figure1}
\end{figure}

The dependence of the actual speed of our code on $N$ is shown in the right panel of Fig. 1.
The code performance (in GFlops) scale as

\begin{equation}
R_{max}=\frac{N(182+29N)}{10^{9}t_{k}}\,,
\end{equation}

\noindent where we counted 182+29\emph{N} operations per single thread and $t_{k}$
is the time (in $sec$) to accomplish a single computational kernel. The sustained performance is more than 11TFlops. This is a very high
value although reached using a hybrid cluster which is small if compared to the top 10 supercomputers in the world.
It is worth to underlining that the larger is the number of stars (until a certain threshold) 
the better is the computational powe of this kind of architecture. Actually, for small $N$
the performance of GPUs are not fully exploited because
the time spent in memory transfers becomes comparable to
that spent in calculating interparticle accelerations. The best GPU performance is achieved when $N$ is large enough that 
all the GPU's CUDA cores are fully loaded.

\section{Conclusions}

Hybrid ( CPU + GPU) cluster architectures are probably the best choices
as a means of investigating the gravitational \emph{N}-body problem. Our next stage
of testing will thus comparing the GPUs from different manufacturers
showing different features. 
Actually, on the market are now found many GPUs apt to perform computations at, nominal, high speed and efficiency, but whose actual suitability to large 
scale physics computations still remains to be checked.

\end{document}